\documentclass[runningheads]{llncs}
\usepackage{graphicx}
\usepackage{booktabs}
\usepackage{url}
\usepackage{adjustbox}

\begin{document}
\title{Generative AI for Medical Imaging: extending the MONAI Framework}

\titlerunning{MONAI Generative Models}

\author{
Walter H. L. Pinaya \inst{1} \and
Mark S. Graham \inst{1} \and
Eric Kerfoot \inst{1} \and
Petru-Daniel Tudosiu\inst{1} \and
Jessica Dafflon\inst{2} \and
Virginia Fernandez\inst{1} \and
Pedro Sanchez\inst{3} \and
Julia Wolleb\inst{4} \and
Pedro F. da Costa\inst{1} \and
Ashay Patel\inst{1} \and
Hyungjin Chung\inst{5} \and
Can Zhao\inst{6} \and
Wei Peng\inst{7} \and
Zelong Liu\inst{8} \and
Xueyan Mei\inst{8} \and
Oeslle Lucena\inst{1} \and
Jong Chul Ye\inst{5} \and
Sotirios A. Tsaftaris\inst{3} \and
Prerna Dogra\inst{6} \and
Andrew Feng\inst{6} \and
Marc Modat\inst{1} \and
Parashkev Nachev\inst{9} \and
Sebastien Ourselin\inst{1} \and
M. Jorge Cardoso\inst{1}
}
\authorrunning{W. Pinaya et al.}
%
\institute{King's College London, London, United Kingdom \and
National Institute of Mental Health, Bethesda, MD, USA \and
The University of Edinburgh, Edinburgh, United Kingdom \and
University of Basel, Allschwil, Switzerland \and
Korea Advanced Institute of Science \& Technology, Daejeon, South Korea \and
NVIDIA Corporation, Santa Clara and Bethesda, USA \and
Stanford University, Stanford, CA, USA \and
Icahn School of Medicine at Mount Sinai, Leon and Norma Hess Center for Science and Medicine, New York, NY, USA \and
University College London, London, United Kingdom}
\maketitle              

\begin{abstract}

Recent advances in generative AI have brought incredible breakthroughs in several areas, including medical imaging. These generative models have tremendous potential not only to help safely share medical data via synthetic datasets but also to perform an array of diverse applications, such as anomaly detection, image-to-image translation, denoising, and MRI reconstruction. However, due to the complexity of these models, their implementation and reproducibility can be difficult. This complexity can hinder progress, act as a use barrier, and dissuade the comparison of new methods with existing works. In this study, we present MONAI Generative Models, a freely available open-source platform that allows researchers and developers to easily train, evaluate, and deploy generative models and related applications. Our platform reproduces state-of-art studies in a standardised way involving different architectures (such as diffusion models, autoregressive transformers, and GANs), and provides pre-trained models for the community. We have implemented these models in a generalisable fashion, illustrating that their results can be extended to 2D or 3D scenarios, including medical images with different modalities (like CT, MRI, and X-Ray data) and from different anatomical areas. Finally, we adopt a modular and extensible approach, ensuring long-term maintainability and the extension of current applications for future features.

\keywords{Generative AI \and Generative Models  \and Medical Imaging \and Deep Learning}
\end{abstract}

\section{Introduction}

In recent years, deep learning has achieved several groundbreaking accomplishments that were made possible through access to vast amounts of data. Since deep neural networks are notoriously data-hungry, they have thrived in fields with large publicly available datasets, such as computer vision, with datasets like ImageNet \cite{deng2009imagenet} and LAION \cite{schuhmann2022laion}, as well as natural language processing (with access to large textual corpora, like the Wikipedia Text Corpus and CommonCrawl\footnote[1]{https://commoncrawl.org/}). Sharing large healthcare datasets is also a crucial component of advancing medical imaging research, but privacy concerns often impede such efforts. The sensitive personal information contained within healthcare data makes it difficult to share openly with researchers without raising ethical and legal concerns. This barrier limits the rate at which cutting-edge machine learning methods can be developed and deployed in the real world, creating a bottleneck in progress \cite{malin2018between,van2014systematic}.

Generative AI has brought incredible breakthroughs and still has immense potential in medical imaging and healthcare. Generative AI refers to a set of artificial intelligence techniques and models designed to learn the underlying patterns and structure of a dataset and generate new data points that plausibly could be part of the original dataset. Specifically, during training, a generative model tries to estimate the probability distribution of data. Assuming that our data points come from an unknown underlying data distribution $x \sim p_{data}(x)$, we use a model $p_\theta(x)$ (that provides a family of parametrised probability distributions) to try to estimate $p_{data}$. Although the data distribution is highly complex for high-dimensional imaging data, these days, we have a  variety of powerful deep generative models that can learn complex data distributions, such as diffusion models \cite{ho2020denoising,rombach2022high}, autoregressive transformers \cite{parmar2018image,esser2021taming}, generative adversarial networks (GANs) \cite{goodfellow2020generative}, and variational autoencoders (VAEs) \cite{kingma2013auto}. 

One of the key benefits of these trained generative models is their ability to create new content that is not limited to the original training dataset. Thus, there is the potential to generate unlimited novel data points just by sampling from the model distribution. This capacity to generate synthetic data has been one of the main goals of generative models in medical imaging, as it presents a viable solution to the problem of sharing data while preserving patient privacy. Data providers can train their models on their own data and share the results in a privacy-preserving manner rather than directly sharing patient data. Recent medical imaging studies have shown a promising ability to generate high-quality medical images (e.g. \cite{korkinof2018high,sun2022hierarchical,peng2022generating,pinaya2022brain,tudosiu2022morphology,dorjsembe2023conditional}). Furthermore, machine learning models trained to perform a downstream task using datasets that also include synthetic data have achieved similar, or even superior performance, compared to models trained with only real data \cite{azizi2023synthetic,fernandez2022can,tian2023stablerep,ktena2023generative}.

Aside from generating synthetic data, generative models have a wide range of other applications in medical imaging. One such application is anomaly detection, where generative models can be utilized to identify anomalies or abnormalities in medical images. This task is particularly useful for diagnosing diseases and detecting potential health risks. Several studies have demonstrated the effectiveness of generative models in this regard (e.g. \cite{sanchez2022healthy,graham2022denoising,wolleb2022diffusion}). Another application is image-to-image translation, where generative models can be trained to transform images from one modality to another. For example, these models can be used to convert CT scans to MRI scans, label maps to MRI images, or MRI scans to anomaly maps \cite{wolleb2022diffusion,fernandez2022can}. Additionally, generative models can be used for image enhancement, where they can learn to improve the quality of medical images without losing important clinical information. This includes image denoising, removing image artifacts, enhancing image resolution, etc.\ ( for example, \cite{xiang2023ddm}). Finally, generative models can be used for image reconstruction, including MRI and CT reconstruction, where they can reconstruct high-quality images from undersampled or noisy data (e.g. \cite{chung2022score,dar2022adaptive}). This has significant implications for medical imaging, enabling faster imaging of the human body. Overall, the potential applications of generative models in medical imaging are vast, and their use will likely continue to grow.

The rising  popularity of medical generative models has sparked a surge in new terms, concepts, techniques, architectures, and metrics \cite{chen2022generative,kazerouni2022diffusion}. However, this field is accompanied by challenges, including the use of various quality evaluation metrics and the complexity of the models. These issues can hinder progress and the implementation of these models. In response, we developed an open-source platform called MONAI Generative Models, which extends the MONAI framework \cite{cardoso2022monai} to ease the development and deployment of generative models in a standardized way. To illustrate the power of our platform, we discuss five experiments where we have used it in a medical imaging context with different modalities (CT, MRI, and X-Ray data) that spans topics such as out-of-distribution detection, image translation, and super-resolution. Our platform has demonstrated its versatility through its application to different modalities and body parts, both in 2D and 3D scenarios, providing an innovative approach to advancing medical imaging.

\section{Features}
In this section, we present the main features included in MONAI Generative models. The Python package and its documentation is available at \url{https://github.com/Project-MONAI/GenerativeModels}.  

\subsection{Models}
Diffusion models have demonstrated remarkable efficacy in medical imaging \cite{kazerouni2023diffusion}. These probabilistic generative models produce images by progressively eliminating noise from an initial Gaussian noise image, denoted as $x_T \sim \mathcal{N}(0, I)$. They are built upon two complementary random processes: the forward process involves the gradual addition of Gaussian noise to a clean image, while the backward process aims to iteratively denoise $x_T$ to obtain a cleaner image at each step. This denoising process is performed by a neural network, denoted as $\theta(x_t, t)$, which predicts the added noise component. Once trained, each step of the backward process involves applying $\theta$ to the current $x_t$ and incorporating a Gaussian noise perturbation to yield a cleaner $x_{t-1}$. One of the goals for MONAI Generative Models was to offer the components to train diffusion model-based methods efficiently. For that, we included the class \texttt{DiffusionModelUNet}, which corresponds to the UNet structure from the diffusion models that have the timesteps conditioning its behaviour via residual connections, as well as spatial transformer layers, which allows for the conditioning of external information into the diffusion models, such as text and numerical scores. Alongside the network architecture, the diffusion model relies on the scheduling of Gaussian noise in each timestep of its sampling process. To facilitate this scheduling process, we have introduced the \texttt{Scheduler} class, which can be specialized into various classes, including \texttt{DDIMScheduler} \cite{song2020denoising} and \texttt{PNDMScheduler} \cite{liu2022pseudo}. Additionally, it is possible to use different profiles for noise, where we created a modular approach that easily extend the current list, comprising of \texttt{linear} \cite{ho2020denoising}, \texttt{scaled linear} \cite{rombach2022high}, and \texttt{cosine} \cite{nichol2021improved} profiles.

Having established the fundamental components required for the diffusion model, our framework extends its support to encompass other diffusion-based models. One notable addition is the Latent Diffusion Model (LDM) \cite{rombach2022high,pinaya2022brain}. These models leverage the latent representation learned by trained compression models to train on higher dimensional 2D or 3D data than traditional diffusion models are capable of. As proposed in \cite{rombach2022high}, this task can be accomplished using an autoencoder with KL-regularization or a Vector-Quantized Autoencoder (VQ-VAE). We have implemented both of these compression models in our package (\texttt{AutoencoderKL} and \texttt{VQVAE} classes), making these models readily interchangeable. Methods based on diffusion models have been used in anomaly detection approaches. One of these corresponds to the work from \cite{wolleb2022diffusionanomaly}, which uses part of the UNet structure to obtain a latent representation of the input images. To support this, we also included the \texttt{DiffusionModelEncoder} class. We have also implemented ControlNets, which have demonstrated remarkable results in image translation tasks \cite{zhang2023adding}. These networks function similarly to Hypernetworks \cite{ha2016hypernetworks}, where an adapter, connected to the original UNet of the diffusion model, influences its behaviour and generates images that closely align with the provided conditioning.

Transformers have demonstrated remarkable performance in generative models, particularly with the autoregressive (decoder-only) format. Autoregressive generative models are one practical approach that provides explicit modelling of the likelihood function. These models define the joint distribution using conditionals over each input feature, given the values of the previous features $p(x)=\prod_{i=1}^{n} p(x_i | x_1, \ldots, x_{i-1})$. For this reason, they are particularly effective on time series and sequential data. One defining characteristic of transformers is their reliance on attention mechanisms \cite{vaswani2017attention}, which capture interactions between inputs irrespective of their relative positions. In each transformer layer, a (self-)attention mechanism is employed, involving three vectors: query, key, and value. Leveraging attention mechanisms, autoregressive transformers excel at predicting the next token in a sequence based on preceding tokens. However, as transformers are designed to work with sequential data, it becomes necessary to transform higher-dimensional data, such as images with 2 or 3 dimensions, into a 1D vector. To address this, we also provide the \texttt{Ordering} class, which offers various methods for defining the ordering of components. These methods include raster scan ordering, S-curve ordering, and random ordering. The \texttt{Ordering} class facilitates the transformation of image data into a sequential format suitable for input to the transformer model. Similar to diffusion models, transformers can be applied to latent representations of a VQ-VAE, where we use the discrete indices of the codebook to define the tokens.

Leveraging on the capacity of GANs, we also included components from the SPatially-Adaptive (DE)normalization (SPADE) study \cite{park2019semantic}. The key innovation of SPADE lies in its ability to incorporate spatially adaptive normalization techniques, which enable the generator to handle variations in object shape and appearance. By using a semantic layout as input, SPADE dynamically adjusts the normalization parameters to match the characteristics of different regions in the generated image, allowing for precise control over the output. This adaptive normalization not only ensures the preservation of fine-grained details, but also enables the generation of diverse and visually appealing images. SPADE has proven effective in various image synthesis tasks, including image-to-image translation for medical imaging \cite{fernandez2022can}. 

In order to incorporate adversarial elements used during the training of SPADE and other networks, we have included the Patch-based discriminator (\texttt{PatchDiscriminator} class) as a supporting component. This discriminator was initially introduced in the Pix2Pix study \cite{isola2017image} and has proven to be effective in various applications. Notably, it is also utilized when training compression models for latent models, such as autoencoders with KL regularization and VQ-VAEs with adversarial components, also known as VQ-GAN \cite{esser2021taming}. In addition to its original implementation, we also included the improved version of the patch-based discriminator introduced in the Pix2PixHD study \cite{wang2018high} (\texttt{MultiScalePatchDiscriminator} class). In this extension, multiple discriminators are employed to assess the realism of images at different resolutions. This multi-resolution approach enhances the discrimination capability of the model by considering image details across different scales. By incorporating the patch-based discriminator and its extensions, our framework enables the integration of adversarial components into various models, which can improve performance and fidelity in generative tasks.

\subsection{Metrics and Losses}

During the development of a generative model, an essential aspect is the ability to quantitatively evaluate the model's performance and the quality of its synthetic samples. Various criteria can be employed to assess the quality of synthetic data. One such criterion is the fidelity of synthetic images, which indicates the degree to which the generated data resembles visually-realistic images. To measure this fidelity, it is common to analyse the similarity between the distributions of synthetic and real data. The Fréchet inception distance (FID) \cite{heusel2017gans} and the Maximum Mean Discrepancy (MMD) \cite{gretton2012kernel} are commonly utilised for this purpose. 

Additionally, the diversity of synthetic data, which signifies how much it covers the full variability of real data, can also be measured. In the pixel or voxel space, the distance between pairs of synthetic images can be calculated for this evaluation. The Multi-Scale Structural Similarity Index Measure (MS-SSIM) \cite{wang2003multiscale} is a metric that incorporates image details at different resolution scales to assess this diversity. These specialised metrics are added to our package adhering to MONAI conventions and can be seamlessly integrated into existing workflows for comprehensive evaluation of generative models.

Besides the metrics, we added new loss functions following the MONAI convention to support the training of generative models. Our framework incorporates spectral losses, which assess the similarity between two images in the frequency domain \cite{dhariwal2020jukebox}. We have also implemented a patch-based adversarial loss, enabling models to utilize discriminator network predictions to guide parameter optimization \cite{isola2017image}. Furthermore, perceptual losses have been integrated, serving as functions that measure the similarity between deep features extracted from two images using a pre-trained network \cite{zhang2018unreasonable,johnson2016perceptual}. Our package offers not only pre-trained networks that have been trained on computer vision datasets such as Imagenet, but also networks pre-trained on datasets specifically designed for medical imaging. For the 2D scenario, we provide networks trained on the RadImageNet dataset \cite{mei2022radimagenet}, a radiology-specific alternative to the ImageNet database, comprising over 1.35 million images. For the 3D scenario, we leverage the MedicalNet \cite{chen2019med3d}, which has been trained using 23 diverse medical 3D datasets. Moreover, considering the computational memory constraints associated with 3D networks, we have implemented a 2.5D approach to compute the perceptual loss. This approach enables the utilization of any available 2D network on a subset of the possible slices in the three possible orientations, effectively balancing computational demands while maintaining the benefits of the perceptual loss calculation.

\section{Experiments and Applications}
In this section, we present experiments and applications intending to demonstrate the capabilities of MONAI Generative Models. The code for all experiments is available at \url{https://github.com/Warvito/generative_monai}.

\subsection{Experiment I: Adaptability to different types of medical image}
One of the goals of our platform is to establish a standardized and unified framework facilitating the development and evaluation of multiple models under consistent conditions. An inherent advantage of extending the components offered by the MONAI platform is its ability to transcend the constraints of 2D scenarios, thereby enabling a seamless application to 3D data. This capacity is particularly valuable given the inherent complexities of training and evaluating models on 3D data. The inherent flexibility to effortlessly transition between different scenarios enables increased adaptability of our models' implementations, facilitating comparative analyses across diverse cases and extending the original scope of their studies. To showcase this attribute, we assessed the generative capacities of one of the state-of-the-art models present in our package, the Latent Diffusion Model, on diverse datasets containing different body parties and modalities.

In this experiment, we trained Latent Diffusion Models in different datasets and body parts to showcase our models' adaptability. We used the following datasets: the MIMIC-CXR dataset \cite{johnson2019mimic} with 96,161 2D chest X-ray images (each with $512 \times 512$ pixels); the CSAW-M dataset \cite{sorkhei2021csaw} with 9,523 2D mammograms (each with $640 \times 512$ pixels); the UK Biobank dataset \cite{sudlow2015uk} with 41,162 3D T1-weighted brain images (each with $160 \times 224 \times 160$ voxels) and a 2D scenario with 360,525 extracted 2D brain slices (each with $160 \times 224$ pixels); the retinal optical coherence tomography (OCT) dataset \cite{kermany2018labeled} with 84,483 2D OCT images (each with $512 \times 512$ pixels).

\begin{figure}[ht!]
\includegraphics[width=\textwidth]{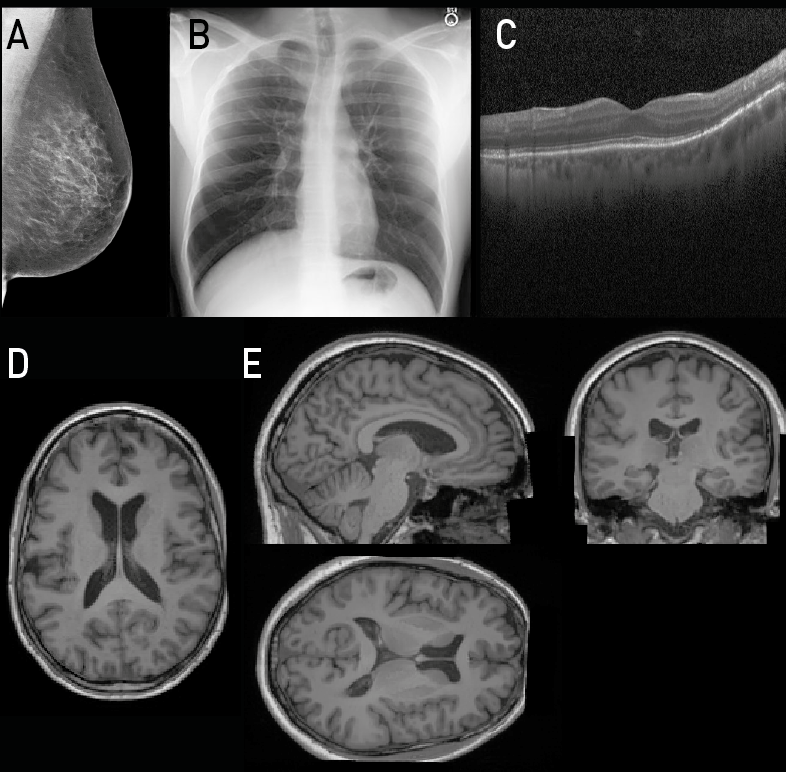}
\caption{Synthetic images from Latent Diffusion Models. A) Mammogram, B) Chest X-Ray, C) Retinal OCT, D) 2D slice from T1-weighted brain image, and E) axial, coronal, and sagittal view of a 3D brain image.} \label{fig:synthetic}
\end{figure}

Table \ref{tab1} presents the performance of the trained models. Since the latent diffusion model contains two models, an autoencoder to learn the latent representation and then a diffusion model to learn a generative model of the latents, we quantify the performance of the autoencoder using the MS-SSIM between the input image and its reconstruction. We evaluate the quality of the samples generated by the diffusion model regarding sample fidelity by measuring the FID between the distribution of synthetic images and our test sets, and the sample diversity by measuring the MS-SSIM between pairs of synthetic images. As shown in Table \ref{tab1} and Figure \ref{fig:synthetic}, our results demonstrate the exceptional performance of generating high-quality images across all evaluated scenarios.

\begin{table}
\caption{Performance of the Latent Diffusion Models on image synthesis.}\label{tab1}
\centering
\begin{tabular}{l c c c c c}
\hline
Dataset & Data Type & Dimensions & FID & MS-SSIM & MS-SSIM Recons.\\
\hline
MIMIC-CXR & 2D Chest X-ray & $512 \times 512$ & 8.8325 & 0.4276 & 0.9834 \\
CSAW-M & 2D Mammography & $640 \times 512$ & 1.9061 & 0.5356 & 0.9789 \\
Retinal OCT & 2D OCT & $512 \times 512$ & 2.2501 & 0.3593 & 0.8966 \\
UK Biobank & 2D Brain Slice MRI & $160 \times 224$ & 2.1986 & 0.5368 & 0.9876 \\
\hline
UK Biobank & 3D Brain MRI & $160\times224\times160$ & 0.0051 & 0.9217 & 0.9820 \\
\hline
\end{tabular}
\end{table}

\begin{figure}[ht!]
\centering
\includegraphics[width=0.6\textwidth]{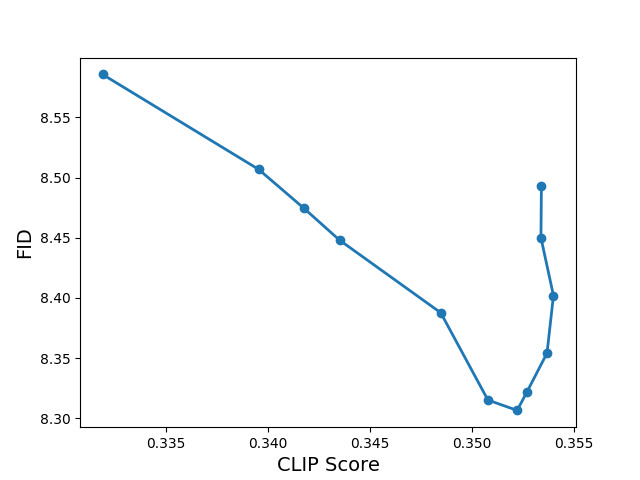}
\caption{CLIP Score vs FID Pareto curve for the LDM trained on chest X-Ray data. We sweep over guidance values of [1, 1.5, 1.75, 2, 3, 4, 5, 6, 7, 8, 9, 10]} \label{fig:clip_score}
\end{figure}

Additionally, MONAI generative models allow us to train LDMs using conditioning variables. In this experiment, we also evaluate how well the models learned the text prompt conditioning used during training by assessing the CLIP Score (more details about the conditioning in Appendix B). By using the BiomedCLIP model \cite{zhang2023large}, we are able to evaluate how close are the image representations of chest X-ray images to the text representations (i.e., the image-text alignment). Using the model trained on the MIMIC-CXR dataset, we generated 1000 images, where we had 125 images from each of the following text prompts: ``atelectasis", ``cardiomegaly", ``no findings", ``edema", ``enlarged cardiomediastinum", ``pleural effusion", ``pneumonia", and ``pneumothorax". This sampling process was performed using the classifier free guidance method \cite{ho2022classifier}, where we can specify the weight for the conditioning, in this experiment we analysed guidance values of [1, , 1.5, 1.75, 2, 3, 4, 5, 6, 7, 8, 9, 10]. Since guidance weight is important to control text alignment, we report the results using the FID vs CLIP score Pareto curve (Figure \ref{fig:clip_score}). As expected, we obtained a high CLIP score, and our curve shows that the CLIP score increases as the guidance value for the conditioning increases (as expected). After a guidance value of 5, only the CLIP score kept increasing while the FID got worse, a similar pattern was also observed in other computer vision studies \cite{saharia2022photorealistic}.

\subsection{Experiment II: Modularity of latent models}

In this experiment, we aim to demonstrate the modularity of components within our latent generative models. We have developed a few generative models that consist of two main components: a compression model and a generative model. The compression model is responsible for learning the latent representation of the data, which is then utilized by the generative model. Our package provides an interface for the interaction between these components to facilitate seamless interchangeability. 

For demonstration, we conducted a comparative analysis of two combinations in this experiment. The first combination involved a VQ-VAE paired with an autoregressive transformer based on the methodology proposed by \cite{esser2021taming}. The second combination involved a Latent Diffusion Model employing the same VQ-VAE as the compression model, following the approach presented by \cite{rombach2022high}. To evaluate their performance, we utilized the MIMIC-CXR dataset from our previous experiment, employing measures such as the quality and diversity of synthetic images. Additionally, we assessed the reconstruction quality of the VQ-VAE using the MS-SSIM metric.

Our VQ-VAE achieved an MS-SSIM score of 0.9689, comparable to the performance observed when using the Autoencoder with KL regularization, as indicated in Table  \ref{tab1}. Furthermore, Table  \ref{tab2} showcases the performance of the transformer and diffusion models.

\begin{table}
\caption{Performance of latent generative models using the same compression model (VQ-VAE).}\label{tab2}
\centering
\begin{tabular}{l c}
\hline
MODEL & FID \\
\hline
VQ-VAE + Transformer & 9.1995 \\
VQ-VAE + Diffusion Model & 8.0457 \\
\hline
\end{tabular}
\end{table}

\subsection{Experiment III: Application to out-of-distribution detection}
Our platform facilitates the application of generative models to many of the downstream tasks they might be used for in medical imaging. Here we demonstrate their use in performing out-of-distribution detection for 3D imaging data. We implement a technique using VQ-VAE + Transformers to obtain image likelihoods, as described in \cite{graham2022transformer,pinaya2022unsupervised}. Models were trained and evaluated on the openly available, 3D Medical Decathlon datasets \cite{antonelli2022medical}, with the BRATs data selected as the in-distribution dataset and all other classes chosen as out-of-distribution data, with all images resized to $128^3$. Models were trained with parameters as described in \cite{graham2022transformer}. The platform's \texttt{VQVAETransformerInferer} class provides methods that facilitate obtaining the likelihoods from latent models. AUC scores for OOD detection on all classes of the Decathlon dataset are shown in Table \ref{tab:anomaly_detection}.

\begin{table}
\caption{Performance of models for out-of-distribution detection, measured as AUC scores. BRATs was used for in-distribution training. }\label{tab:anomaly_detection}
\centering
\resizebox{\textwidth}{!}{
  \begin{tabular}{l cccccccc }
  \toprule
  &\multicolumn{8}{c}{\bfseries OOD Dataset}\\

  & Liver CT & Hippocampal MR & Prostate MR & Lung CT & Pancreas CT & Hepatic CT & Spleen CT & Colon CT\\
  \midrule
Transformer & 1.0 & 1.0 & 1.0 & 1.0 & 1.0 & 1.0 & 1.0 & 1.0\\

\bottomrule
\end{tabular}
}
\end{table}

\subsection{Experiment IV: Application to Image Translation}
\begin{figure}[ht!]
\includegraphics[width=\textwidth]{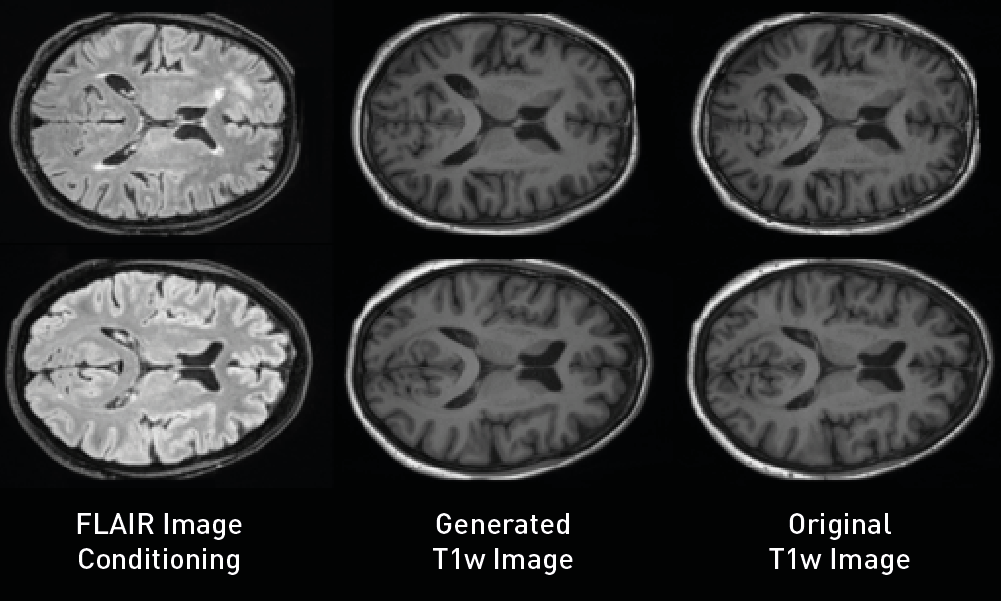}
\caption{Image translation from FLAIR to T1-weighted MRI images with ControlNets.} \label{fig:img2img}
\end{figure}

As previously mentioned, another type of application that we aim to support is image-to-image tasks. This experiment will present how ControlNets performs in an image translation task. ControlNets are a neural network that significantly enhances the controllability and customization of diffusion models. They work like lightweight adapters that control the behaviour of a pre-trained network. In this experiment, we used the model trained on the UK Biobank 2D dataset (Experiment I) to train a ControlNet capable of transforming 2D FLAIR slices into the T1-weighted images that the diffusion model creates. To evaluate the performance of the model, we used the Mean Absolute Error, the Peak Signal Noise Ratio, and the MS-SSIM between the synthetic 2D image and the true T1-weighted image. Our model was capable of generating the T1-weight images with high fidelity on the conditioning FLAIR image (Figure \ref{fig:img2img}), obtaining a performance of PSNR=$26.2458 \pm 1.0092$, MAE=$0.02632 \pm 0.0036$ and MS-SSIM=$0.9526 \pm 0.0111$.

\subsection{Experiment V: Application to image super-resolution}
Generative models have proven highly effective in a wide range of tasks, including super-resolution. This task is not only valuable for improving the quality of low-resolution images but also enables image generation through the use of cascaded models \cite{ho2022cascaded}. The cascade approach involves training a generative model to generate images with lower resolution, followed by a secondary model that focuses on achieving the desired higher resolution. This approach is particularly advantageous for 3D models, allowing training with a smaller memory footprint.

In our study, we explored the application of generative models for super-resolution using the Stable Diffusion 2.0 Upscaler approach. Leveraging 3D data from the UK Biobank, we initially trained a Latent Diffusion Model on a version of the dataset with a resolution of 2 $mm^3$ ($80\times112\times80$ voxels). To further enhance the results, we trained another LDM on patches of $32\times32\times32$ voxels.

To incorporate the super-resolution technique, we concatenated the low-resolution image as input to the diffusion model and added Gaussian noise. The diffusion model was conditioned based on the noise level, allowing for better performance. By iteratively applying super-resolution techniques to tiles of the generated image, we achieved significant improvements in image quality and resolution.

\begin{figure}[ht!]
\includegraphics[width=\textwidth]{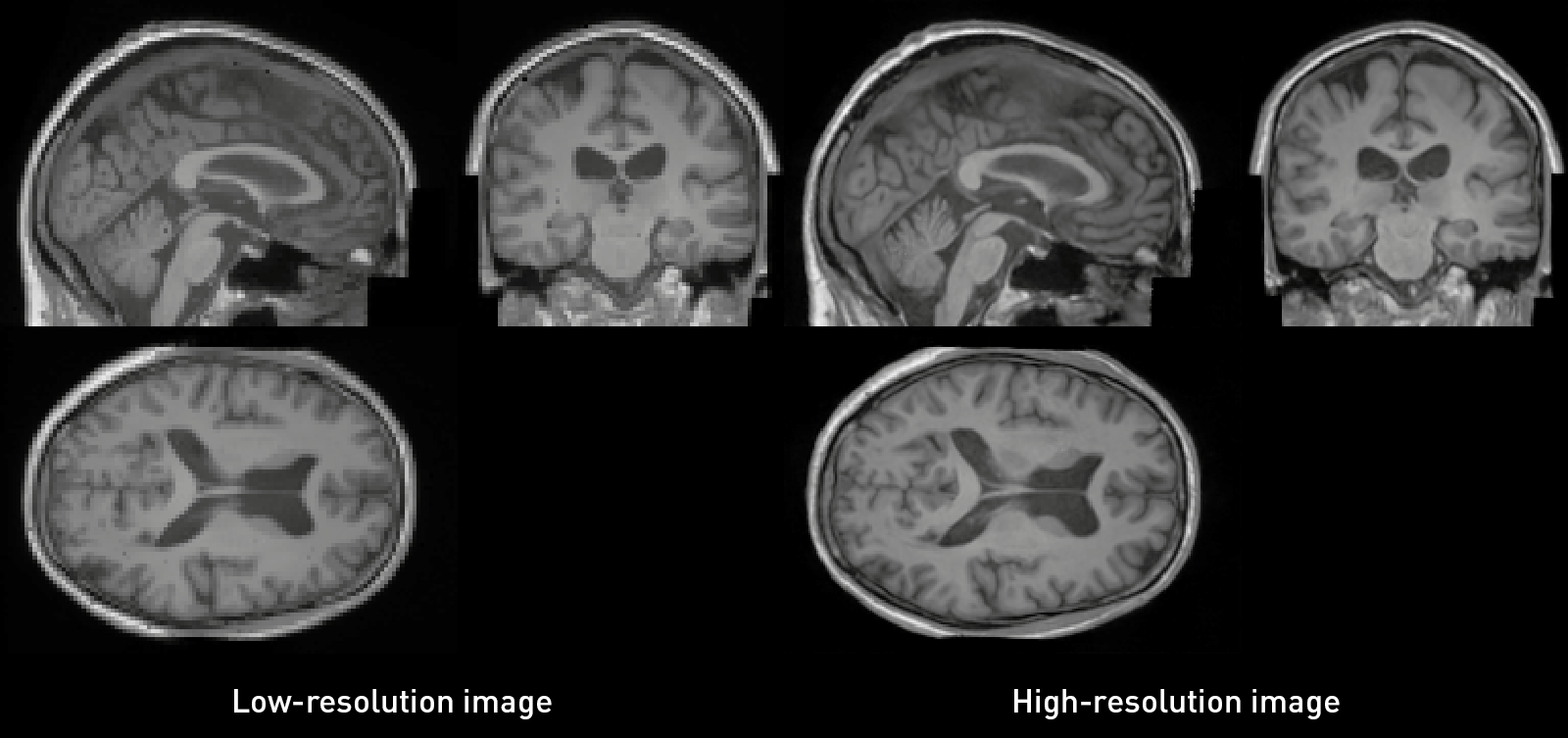}
\caption{3D super-resolution with Stable Diffusion Upscaler method.} \label{fig:superres}
\end{figure}

Our study demonstrates the effectiveness of generative models in super-resolution tasks, particularly for 3D models (Figure \ref{fig:superres}). The low-resolution model was evaluated against a downsampled test set, achieving an FID=$0.0009$, MS-SSIM=$0.7818$ for diversity, and MS-SSIM=$0.9970$ for autoencoder reconstruction. We generated high-resolution synthetic images using the conditioned upscaler model with a noise level of 1. Comparing them to the high-resolution test set, we observed an FID of 0.0024 and MS-SSIM of 0.9141, showing a higher fidelity and similar diversity when comparing this cascaded approach to a single-step generation (from Experiment I). This improvement was possible due to the increased number of parameters in this experiment's networks while using the same amount of computational resources to train them. On the other hand, image generation takes significantly more time, taking about 22 seconds to generate an image using the single-stage approach and about 13 minutes to generate it using the cascaded approach (using an NVIDIA TITAN RTX and DDIMScheduler with 200 timesteps). 

In the final part of this experiment, we assessed the performance of our model in super-resolving images. To accomplish this, we upscaled the low-resolution images from the test set and compared them to the corresponding ground truth images. Our model demonstrated excellent results, achieving a PSNR = $29.8042 \pm 0.4173$, MAE=$0.0181 \pm 0.0009$ and MS-SSIM=$0.9806 \pm 0.0017$. These evaluation metrics affirm the high-quality super-resolution capabilities of our model, further validating its effectiveness in enhancing image resolution. These results highlight the potential of generative models for enhancing image quality and generating high-resolution images.

\section{Conclusions}
In this study, we have presented MONAI Generative Models, an extension of the widely used open-source MONAI platform, designed to assist users in developing and deploying various generative models. By leveraging the modularity and adaptability offered by MONAI Generative Models, researchers can significantly reduce the time and effort required when applying these models, not only for image synthesis but also for diverse applications. Looking ahead, our future work will focus on incorporating more recent models to enable easier model comparison, as well as enhancing support for other applications, such as MRI reconstruction. These advancements will undoubtedly contribute to further advancements in the field of medical generative models and their practical implementations.

\bibliographystyle{splncs04}
\bibliography{mybibliography.bib}

\newpage
\section*{Appendix A - Additional Samples}

\begin{figure}[ht!]
\centering
\includegraphics[width=\textwidth]{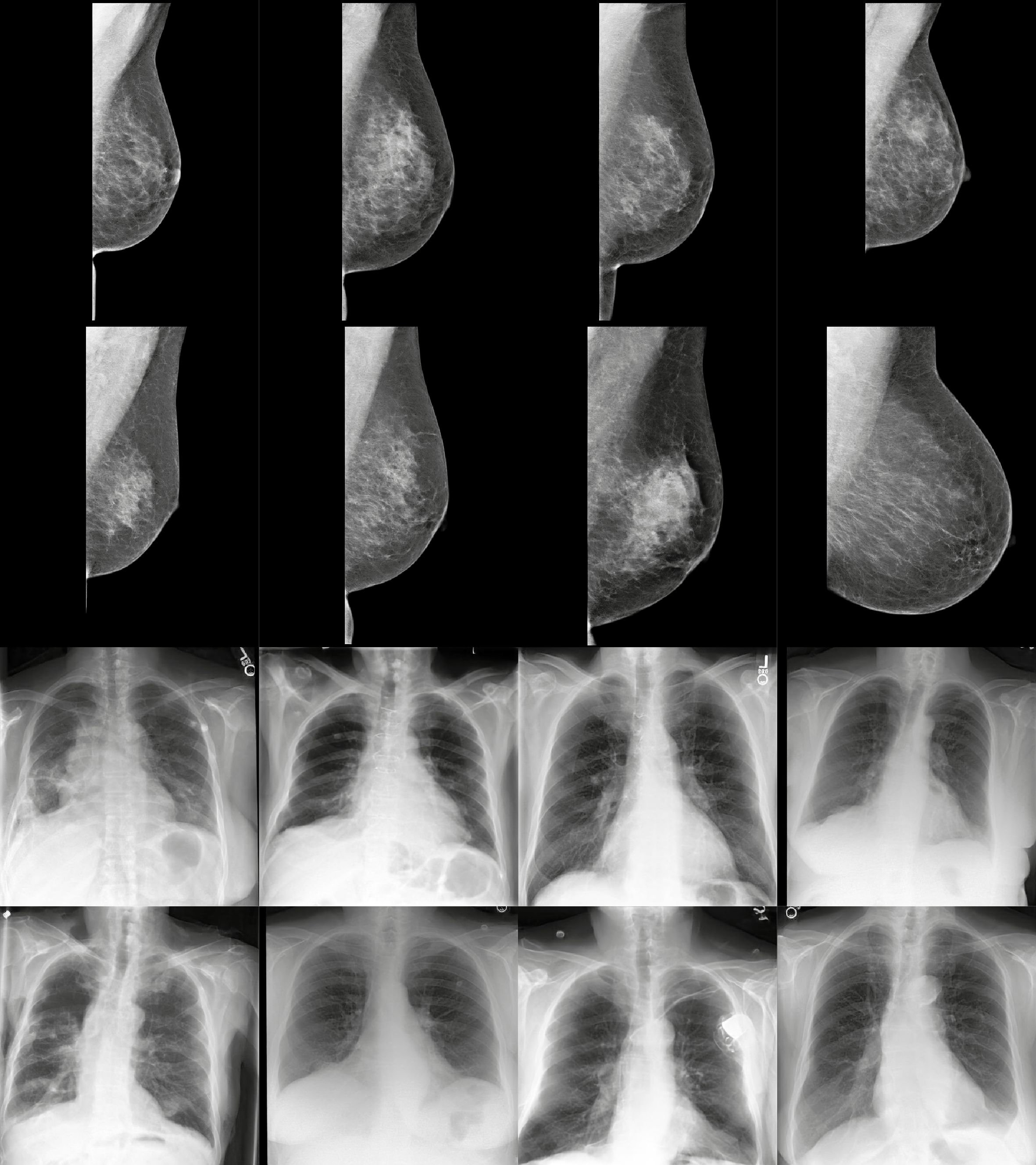}
\caption{Random samples from LDMs trained on the CSAW-M and MIMIC-CXR datasets. Sampled with 200 DDIM steps.} \label{fig:supp1}
\end{figure}

\begin{figure}[h!]
\centering
\includegraphics[width=0.90\textwidth]{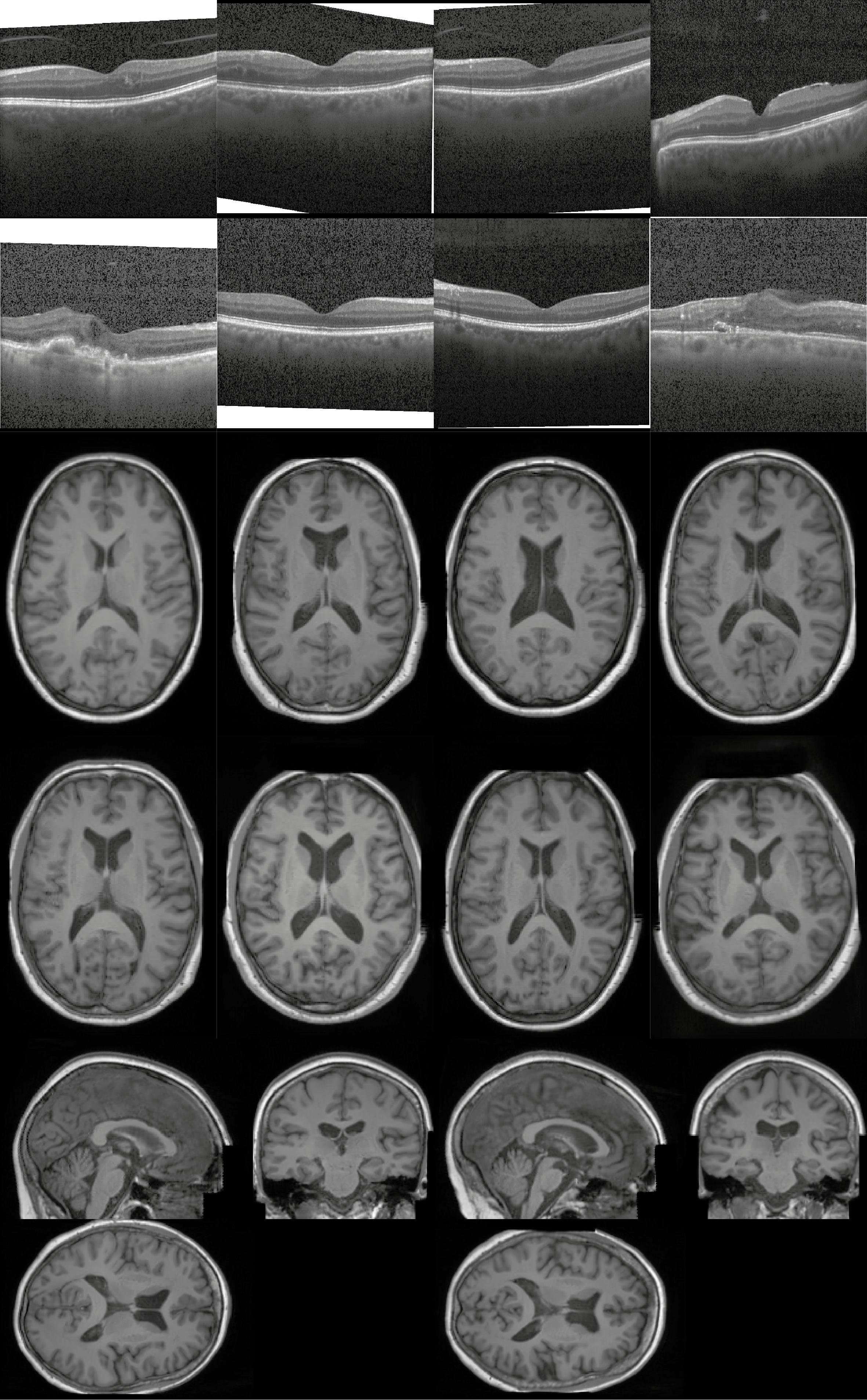}
\caption{Random samples from LDMs trained on the Retinal OCT, 2D UK Biobank, and 3D UK Biobank datasets. Sampled with 200 DDIM steps.} \label{fig:supp2}
\end{figure}

\begin{figure}[ht!]
\centering
\includegraphics[width=0.95\textwidth]{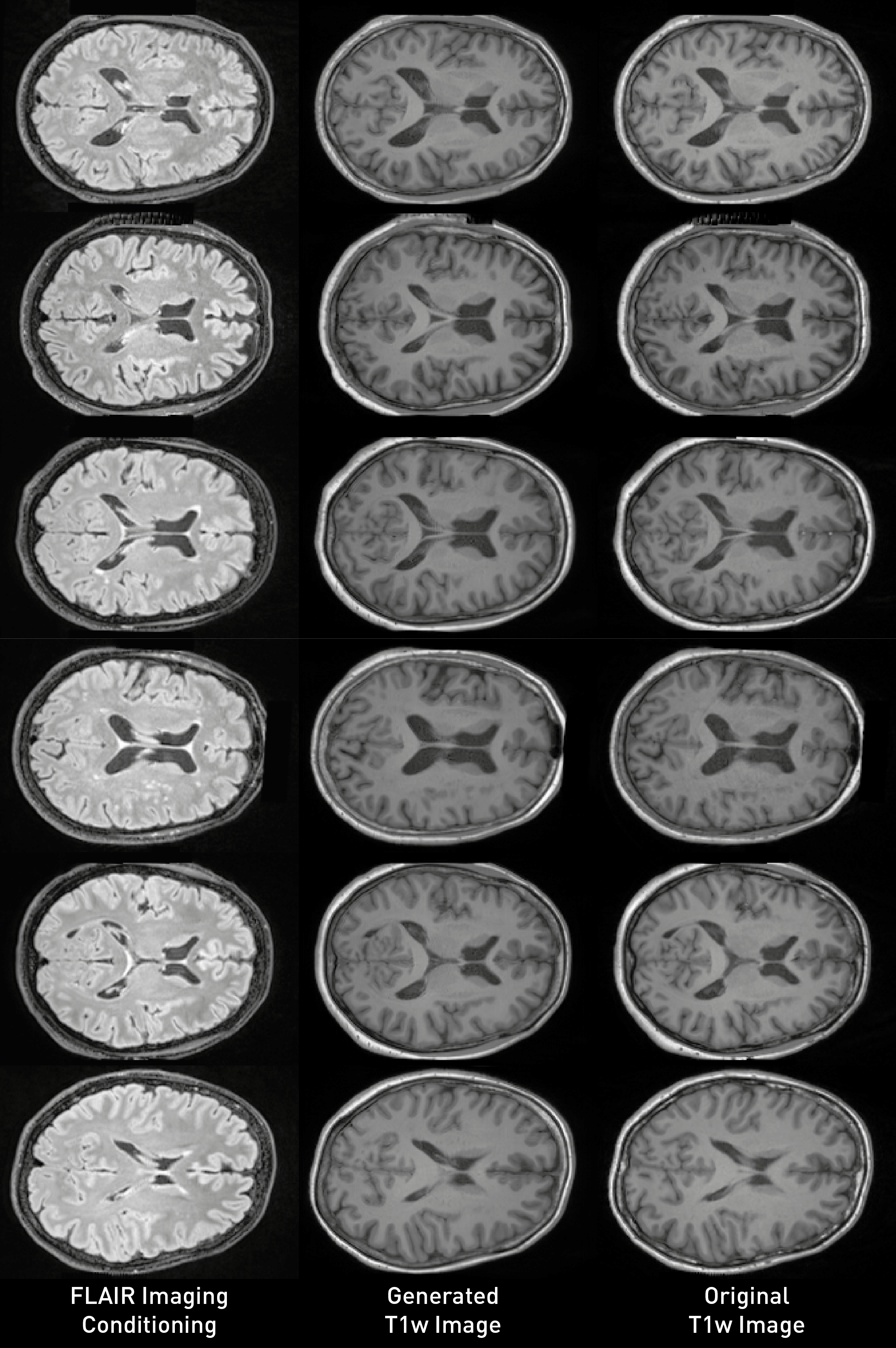}
\caption{Random samples from ControlNets trained on the UK Biobank dataset.} \label{fig:supp3}
\end{figure}

\begin{figure}[ht!]
\includegraphics[width=0.95\textwidth]{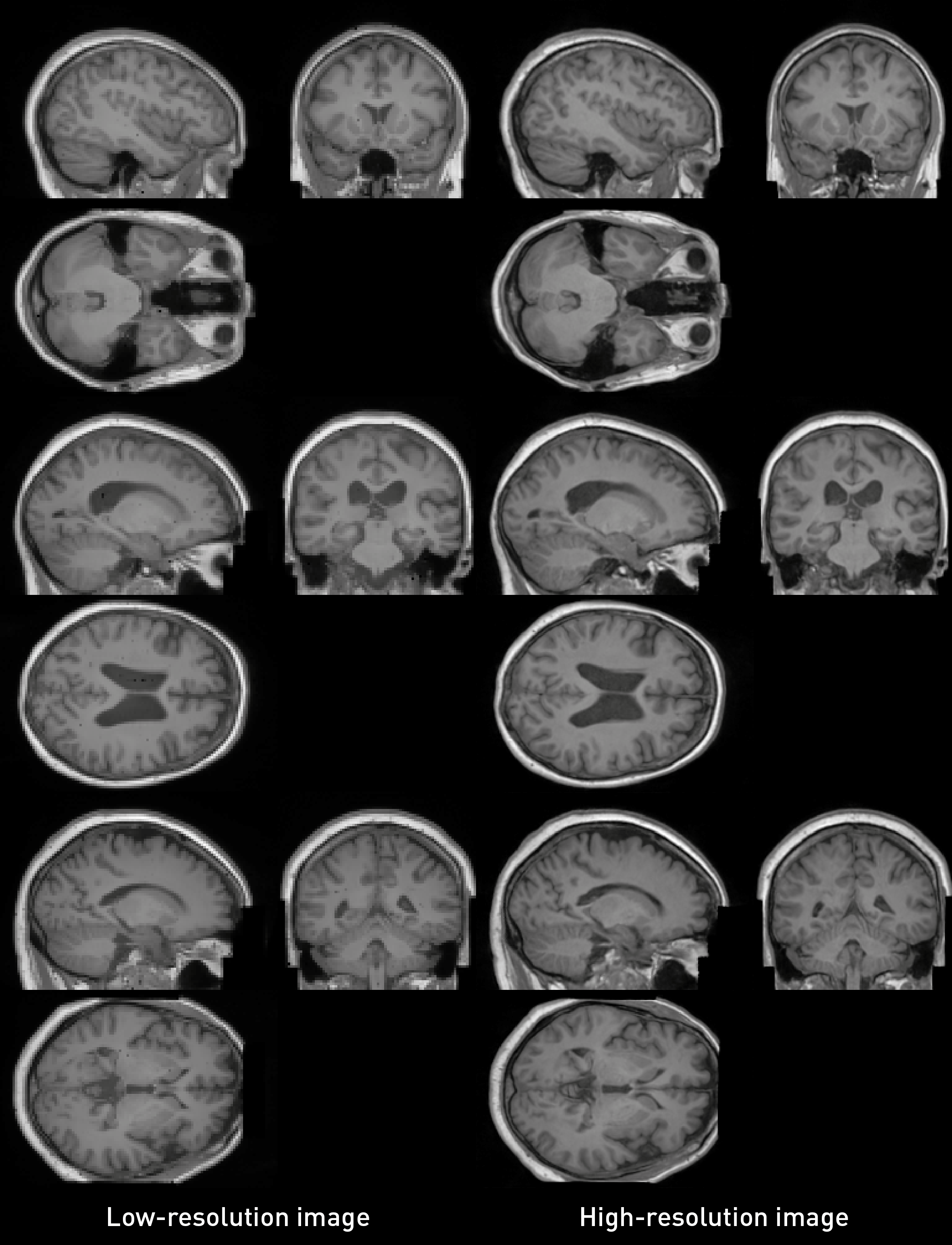}
\caption{Random samples 
obtained from the Latent Diffusion Model trained on the 3D low resolution UK Biobank dataset and the same ones super-resolved.} \label{fig:supp4}
\end{figure}

\newpage
\clearpage
\section*{Appendix B - Hyperparameters and Implementation Details}
We provide an overview of the hyperparameters of all trained models in Experiment I (Table \ref{tab_supp1} and \ref{tab_supp2}). In all compression models, we did not include attention layers. All 2D compression models had the same configuration using an L1 loss, an adversarial component weight = 0.005, a perceptual loss weight = 0.002 (with the perceptual loss computed using the \texttt{``squeeze"} pre-trained network), and a KL regularization weight 1e-8. We trained these models using an Adam Optimizer with a learning rate = 0.00001 and a discriminator learning rate = 0.00005. For the 3D scenario, we used a learning rate = 0.00005 and a discriminator learning rate = 0.0001. For the computation of the perceptual loss, we used the 2.5D approach using a  \texttt{``squeeze"} pre-trained network on 25\% of the slices.
  
\begin{table}
\caption{Hyperparameters for the compression models (Autoencoder with KL regularization) from Experiment I.}\label{tab_supp1}
\centering
\begin{tabular}{l c c}
\hline
 & 2D networks & 3D network \\
\hline
Compression Factor & 8 & 8 \\
\# Channels & [128, 256, 256, 512] & [32, 64, 128, 128]  \\
\# Latent Channels & 4 & 3 \\
\# Residual Blocks & 2 & 2  \\
\hline
Patch Discriminator &  &  \\
\hline
\# Channels & 64 & 96 \\
\# Layers & 3 & 3 \\
\hline
\end{tabular}
\end{table}

Table \ref{tab_supp2} show the hyperparameters used on the diffusion models from Experiment I. The diffusion models were trained using an AdamW Optimizer with a learning rate = 0.000025. For all models, we used a ``scaled linear" schedule (originally from \cite{rombach2022high}) with 1000 timesteps, beta range [0.0015, 0.0205], and a prediction type = \texttt{v\_prediction}.

\begin{table}
\caption{Hyperparameters for the diffusion models from Experiment I.}\label{tab_supp2}
\centering
\begin{adjustbox}{max width=\textwidth}
\begin{tabular}{l c c}
\hline
 & 2D networks & 3D networks \\
\hline
\# Channels & [320, 640, 1280, 1280] & [256, 512, 768]  \\
Attention levels & [False, True, True, True] & [False, True, True] \\
Cross-Attention dimension & 1024 & 1024  \\
Attention Head \# of Channels & [0, 128, 128, 128] & [0, 512, 768] \\
\hline
\end{tabular}
\end{adjustbox}
\end{table}

We conditioned all diffusion models on a text prompt during the training. For that, we used the tokenizer and the text encoder from the Stable Diffusion 2.1, available via the \texttt{transformers} package\footnote{https://huggingface.co/docs/transformers/index}. For the UK Biobank dataset, we used the text ``T1-weighted image of a brain." for all images. For the Retinal OCT, we used the following text according to the image label: ``Choroidal Neovascularization", ``Diabetic Macular Edema", ``Drusen", ``Normal retina". For the CSAW-M dataset, we used a different text prompt according to the masking level of cancer: ``Low masking level" ($<=2$), ``Medium masking level" ($>2$ and $<=6$), and ``High masking level" ($>6$). For the MIMIC-CXR dataset, we used up to 5 random sentences sampled from the radiological report of each image as the text prompt. In all models, we trained with a 10\% chance of replacing the text conditioning with an empty string (``"). This allows us to sample models using the classifier free guidance approach \cite{ho2022classifier}. To compute the models' performance (i.e.\ FID and MS-SSIM), we sampled 1000 images in an unconditioned way (using ``") to obtain samples from across the whole data distribution.

For Experiment II, we trained a VQ-GAN using an Adam optimizer with a learning rate = 0.00005 and a discriminator learning rate = 0.0001. We used similar weights to Experiment I when computing the adversarial and perceptual components. We used a network with 64, 128, and 128 channels across the different levels, where we had two residual layers per level. We used 3 channels for the latent space and 512 different elements for the codebook. For the diffusion model, we used 256, 512, and 768 channels, with attention layers in the deepest two layers, each one using just a single attention head. The other parameters are similar to the 2D models used in Experiment I. For the transformer, we trained it using the cross entropy loss function and an AdamW optimizer with a learning rate =  0.000025. The transformer had 513 number of tokens, a maximum sequence of 4096, an internal layer size of 64, and a depth of 8 layers, each with eight attention heads.

For Experiment IV, we used the LDM trained on 2D brain data from Experiment I. We used the same parameters to train the Controlnet, we had the conditioning image processed by a convolutional network with 64, 128, 128, and 256 channels and downsampling layers between them.

We provide an overview of the hyperparameters of all trained models in Experiment V (Table \ref{tab_supp3} and \ref{tab_supp4}). All other details are similar to the ones used in Experiment I.

\begin{table}
\caption{Hyperparameters for the compression models (Autoencoder with KL regularization) from Experiment V.}\label{tab_supp3}
\centering
\begin{tabular}{l c c}
\hline
 & Low resolution LDM & Upscaler LDM \\
\hline
Compression Factor & 4 & 2 \\
\# Channels & [128, 256, 256] & [128, 256]  \\
\# Latent Channels & 4 & 4 \\
\# Residual Blocks & 2 & 2  \\
\hline
Patch Discriminator &  &  \\
\hline
\# Channels & 64 & 64 \\
\# Layers & 3 & 3 \\
\hline
\end{tabular}
\end{table}

\begin{table}
\caption{Hyperparameters for the diffusion models from Experiment V.}\label{tab_supp4}
\centering
\begin{adjustbox}{max width=\textwidth}
\begin{tabular}{l c c}
\hline
 & Low resolution LDM & Upscaler LDM \\
\hline
\# Channels & [512, 512, 768] & [256, 256, 512]  \\
Attention levels & [False, False, True] & [False, False, True] \\
Cross-Attention dimension & 1024 & 1024  \\
Attention Head \# of Channels & [0, 0, 768] & [0, 0, 512] \\
\hline
\end{tabular}
\end{adjustbox}
\end{table}

\end{document}